\documentclass{article} 
\usepackage{arxiv, times}

\usepackage{amsmath,amsfonts,bm}

\def\eqref#1{equation~\ref{#1}}

\def\1{\bm{1}}

\def\vh{{\bm{h}}}

\def\vo{{\bm{o}}}
\def\vp{{\bm{p}}}

\def\vw{{\bm{w}}}
\def\vx{{\bm{x}}}
\def\vy{{\bm{y}}}
\def\vz{{\bm{z}}}

\def\mU{{\bm{U}}}
\def\mV{{\bm{V}}}
\def\mW{{\bm{W}}}

\DeclareMathAlphabet{\mathsfit}{\encodingdefault}{\sfdefault}{m}{sl}
\SetMathAlphabet{\mathsfit}{bold}{\encodingdefault}{\sfdefault}{bx}{n}

\def\sR{{\mathbb{R}}}

\def\sZ{{\mathbb{Z}}}

\newcommand{\softmax}{\mathrm{softmax}}
\newcommand{\sigmoid}{\sigma}

\usepackage{url}            
\usepackage{booktabs}       
\usepackage{amsfonts}       
\usepackage{nicefrac}       
\usepackage{microtype}      
\usepackage{graphicx}
\usepackage{amsmath}
\usepackage{mathtools}
\usepackage{adjustbox}
\usepackage{subfigure}
\usepackage{varwidth}
\usepackage{capt-of}
\usepackage[table,xcdraw]{xcolor}

\usepackage{todonotes}

\title{META$^\mathbf{2}$: Memory-efficient taxonomic classification and abundance estimation for metagenomics with deep learning}

\author{
Andreas Georgiou \\
Department of Computer Science\\
ETH Zurich \\
Zurich, Switzerland \\
\texttt{geandrea@ethz.ch} \\
\And
Vincent Fortuin \thanks{Equal contribution} \\
Department of Computer Science\\
ETH Zurich\\
Zurich, Switzerland \\
\texttt{fortuin@inf.ethz.ch} \\
\And
Harun Mustafa $^*$ \\
Department of Computer Science\\
ETH Zurich\\
Zurich, Switzerland \\
\texttt{harun.mustafa@inf.ethz.ch} \\
\And
Gunnar R\"atsch \\
Department of Computer Science\\
ETH Zurich\\
Zurich, Switzerland \\
\texttt{raetsch@inf.ethz.ch}
}

\begin{document}

\maketitle

\begin{abstract}
Metagenomic studies have increasingly utilized sequencing technologies in order to analyze DNA fragments found in environmental samples.
One important step in this analysis is the taxonomic classification of the DNA fragments. Conventional read classification methods require large databases and vast amounts of memory to run, with recent deep learning methods suffering from very large model sizes. We therefore aim to develop a more memory-efficient technique for taxonomic classification. A task of particular interest is abundance estimation in metagenomic samples. Current attempts rely on classifying single DNA reads independently from each other and are therefore agnostic to co-occurence patterns between taxa. In this work, we also attempt to take these patterns into account. We develop a novel memory-efficient read classification technique, combining deep learning and locality-sensitive hashing. We show that this approach outperforms conventional mapping-based and other deep learning methods for single-read taxonomic classification when restricting all methods to a fixed memory footprint. Moreover, we formulate the task of abundance estimation as a Multiple Instance Learning (MIL) problem and we extend current deep learning architectures with two different types of permutation-invariant MIL pooling layers: a) deepsets and b) attention-based pooling. We illustrate that our architectures can exploit the co-occurrence of species in metagenomic read sets and outperform the single-read architectures in predicting the distribution over taxa at higher taxonomic ranks.
\end{abstract}

\section{Introduction}
\label{sec:intro}

Over the last decades, advancements in sequencing technology have led to a rapid decrease in the cost of genome sequencing~\cite{sequencingcosts}, while the amount of sequencing data being generated has vastly increased. This is attributable to the fact that genome sequencing is a tool of utmost importance for a variety of fields, such as biology and medicine, where it is used to identify changes in genes or aid in the discovery of potential drugs~\cite{HMP, MetaHIT}. Metagenomics is a subfield of biology concerned with the study of genetic material extracted directly from an environmental sample~\cite{metasub2016metagenomics, howe2014tackling}.
Significant efforts have been carried out by projects such as the Human Microbiome Project (HMP)~\cite{HMP} and the Metagenomics of the Human Intestinal Tract (MetaHIT) project~\cite{MetaHIT} in order to understand how the human microbiome can have an effect on human health. An important step in this process is the classification of DNA fragments into various groups at different taxonomic ranks, typically referencing the NCBI Taxonomy for the relationships between the samples against which fragments are matched~\cite{NCBIRef}. In this taxonomy, organisms are assigned taxonomic labels and are thus placed on the taxonomic tree. Each level of the tree represents a different taxonomic rank, with finer ranks such as \textit{species} and \textit{genus} being close to the leaf nodes and coarser ranks such as \textit{phylum} and \textit{class} closer to the root.

We consider the problem of metagenomic classification, where each individual read is assigned a label or multiple labels corresponding to its taxon at each taxonomic rank. One could simply identify the taxon at the finest level of the taxonomy and then extract the taxa at all levels above by following the path to the root. The problem with this approach is that for certain reads, we might not be able to accurately identify the species of the host organism, but nevertheless be interested in coarser taxonomic ranks. This can apply in cases where little relevant reference data is available for a sequencing data set, such as deep sea metagenomics data~\cite{tully2018reconstruction} or public transit metagenomics~\cite{afshinnekoo2015geospatial, danko2019global}, so an accurate prediction at higher taxonomic ranks may be more informative for downstream analysis~\cite{rojas2019genet}.
This happens especially when reads come from regions of the genome that are conserved across species, but may be divergent at higher taxonomic ranks. Furthermore, in many cases we are also interested in the distribution of organisms in an environmental sample alongside the classifications of individual fragments.

The closely related task of sample discovery, in which the closest match to a query sequence is found in an annotated database of reference sequences, has drawn considerable attention in recent years~\cite{muggli2017succinct, garrison2018variation, approxmatching_mantis, approxmatching1, bradley2019ultrafast, bingmann2019cobs, mustafa2019dynamic, karasikov2019sparse}. While these methods may be used as a back end for classification given metadata associating each sample to its respective taxonomic ranks, this usually comes at the expense of large database sizes. To overcome this, these tools provide various means to trade off accuracy for representation size. Exact methods, for example, may allow for the use of more highly compressed, but slower data structures~\cite{mustafa2019dynamic,almodaresi2019efficient}, or for the subsampling of data at the expense of greater false negatives~\cite{approxmatching_mantis}. On the other hand, approximate methods trade off smaller representation size for increased false positives~\cite{approxmatching1,bradley2019ultrafast, bingmann2019cobs}.

Alongside these approaches, deep learning has also shown great promise and has become increasingly prevalent for approaching biological classification tasks. In recent years, we have seen various attempts of using deep learning to solve tasks such as variant calling~\cite{deepSNP} or the discovery of DNA-binding motifs~\cite{DeepGenomicsPrimer}. These methods even outperform more classical approaches, despite the relative lack of biological prior knowledge incorporated into those models.

In this work, we develop new memory-efficient deep learning methods for taxonomic classification. Furthermore, we formulate the task of abundance estimation as an instance of Multiple Instance Learning (MIL). MIL is a specific framework of supervised learning approaches. In contrast to the traditional supervised learning task, where the goal is to predict a value or class for each sample, in MIL, given a set of samples, the goal is to assign a value to the whole set. A set of items is called a \textit{bag}, whereas each individual item in the bag is called an \textit{instance}. In other words, a bag of instances is considered to be one data point~\cite{MILDefinition}. More formally, a bag is a function $\mathbf{B}: \mathcal{X} \rightarrow \mathbb{N}$ where $\mathcal{X}$ is the space of instances. Given an instance $x \in \mathcal{X}$, $\mathbf{B}(x)$ counts the number of occurrences of $x$ in the bag $\mathbf{B}$. Let $\mathcal{B}$ be the class of such bag functions. Then the goal of a MIL model is to learn a bag-level concept $c: \mathcal{B} \rightarrow \mathcal{Y}$ where $\mathcal{Y}$ is the space of our target variable.

In the context of metagenomic classification, we consider the instances to be DNA reads. Our goal is to directly predict the distribution over a given set of taxonomic ranks in the read set (the bag). So for each taxon, our output is a real number in $[0, 1]$, denoting the portion of the reads in the read set that originated from that particular species. The motivation for this is that in a realistic set of reads, certain organisms tend to appear together. It might thus be possible to exploit the co-occurrence of organisms to gain better accuracy~\cite{MILSurvey}.

Our main contributions are:

\begin{itemize}
    \item A new method for learning sequence embeddings based on locality-sensitive hashing (LSH) with reduced memory requirements.
    \item A new memory-efficient deep learning method for taxonomic classification.
    \item A novel machine learning model for predicting the distribution over taxa in a read set, combining state-of-the-art deep DNA classification models with read-set-level aggregation in a multiple instance learning (MIL) setting.
    \item A thorough empirical assessment of our proposed single-read classification models, showing significant memory reduction and superior performance when compared to equivalent mapping-based methods.
    \item A thorough empirical assessment of our proposed MIL abundance estimation models showing comparable performance in predicting the distributions of higher level taxa from read sets with significantly lower memory requirements.
\end{itemize}

In the rest of this paper, we give an overview of previous related work in Section~\ref{sec:relwork}, describe our data generation method and machine learning models in Section~\ref{sec:modelsandmethods} and analyse the results of our experiments in Section~\ref{sec:results}.
An overview of our proposed architectures is depicted in Figures~\ref{fig:archs} and \ref{fig:singlearchs}.

\section{Related Work}
\label{sec:relwork}
To solve the problem of metagenomic classification, traditional methods rely on the analysis of read mappings to classify each DNA fragment. Given a DNA read, one first needs to match $k$-mers to a large database of reference genomes to detect candidate segments of target genomes, followed by
approximate string matching techniques to match the string to these candidate segments.
A well-known and widely used tool that uses alignment is BLAST, which is a general heuristic tool for aligning genomic sequences. Results from these matches can then be used to compute a taxonomic classification using MEGAN~\cite{huson2007megan}.
Other mapping-based tools specifically designed for metagenomics include Centrifuge~\cite{centrifuge}, Kraken~\cite{wood2014kraken}, and MetaPhlAn~\cite{segata2012metagenomic}. These methods make trade-offs of sensitivity for scalability. For example, BLAST is highly sensitive, but not scalable to databases of unassembled sequencing data, while more approximate methods like Kraken are well suited for such large databases. In particular, the accuracy of Kraken can be tailored to the desired size of the final database by selectively discarding the stored $k$-mers, at the expense of an increased number of false negatives. In additional, tools such as Bracken have been developed for estimating the abundances of taxonomic ranks by post-processing read classifications~\cite{lu2017bracken}. Moreover, recent deep learning approaches have outperformed these methods in classifying very long reads with high error-rates~\cite{rojas2019genet}.

Most of the previous attempts using machine learning focused on 16S rRNA sequences due to their high sequence conservation across a wide range of species. An example is the RDP (Ribosomal Database Project) classifier  which uses a Naive Bayes classifier to classify 16S rRNA sequences~\cite{rRNAbayes}. The disadvantage of this method is the loss of positional information due to the encoding of the sequence as a \emph{bag} of 8-letter words. However, the generalizability of this model to sequencing data drawn from other genomic regions is unclear. Similarly, \cite{la2015probabilistic} use probabilistic topic modeling in order to classify 16S rRNA sequences in the taxonomic ranks from phylum to family. Another interesting approach is taken by \cite{brady2009phymm} which uses Markov models to classify DNA reads and can even be combined with alignment methods to increase performance. In addition, \cite{busia2019deep} use a CNN architecture to classify 16S sequences, while other approaches also proposed to use recurrent neural networks on sequences~\cite{ganscha2018supervised}. Other machine learning approaches include \cite{menegaux2019continuous} who learn low-dimensional representations of reads based on their $k$-mer content and \cite{luo2019metagenomic} that combine locality sensitive hashing based on LDPC codes with support vector machines.

More recent attempts for solving the general metagenomic classification problem focus on using deep learning to tackle it as a supervised classification task. Two examples of such attempts are \textit{GeNet}~\cite{rojas2019genet}, which attempts to leverage the hierarchical nature of taxonomic classification, and \textit{DeepMicrobes}~\cite{deepmicrobes}, which first learns embeddings of $k$-mers and subsequently uses those to classify each read. We use \textit{GeNet} and a simplified version of \textit{DeepMicrobes} as baselines and explain them in more detail in Section~\ref{sec:modelsandmethods}. These methods however require significant amount of GPU memory to achieve high classification accuracy. \cite{deepmicrobes}, for example, can reach memory requirements of upto 50 GB when large $k$-mers need to be encoded. Numerous different methods have been proposed for the reduction of such large embeddings, including feature hashing \cite{weinberger2009feature} and hash embeddings \cite{svenstrup2017hash}.

\section{Models and Methods}
\label{sec:modelsandmethods}
For our experiments, we compare against both mapping-based and alignment-free methods such as \textit{Centrifuge}~\cite{centrifuge} and \textit{Kraken}~\cite{wood2014kraken}, and the more recent deep learning techniques \textit{GeNet}~\cite{rojas2019genet} and \textit{EmbedPool}~\cite{deepmicrobes}. Furthermore, we tackle the problem of abundance estimation by adding a multiple instance learning (MIL) pooling layer to our models, in order to capture interactions between reads and directly predict the distribution over species in metagenomic samples.

\subsection{Data set generation}
\label{sec:datagen}

For training, validation, and evaluation of the deep learning models, we use synthetic reads generated from bacterial genomes from the NCBI RefSeq database~\cite{NCBIRef}, from which we use a subset of $3\,332$ genomes comprising $1\,862$ species similar to the data set used in \cite{rojas2019genet}. We use NCBI's \textit{Entrez} tool~\cite{schuler199610}, to download the genomes and the taxonomic data. The number of taxa in each taxonomic rank is summarized in Table~\ref{tab:taxons} in Appendix~\ref{app:supp_results}. The full list of accession numbers for the genomes used is included in our GitHub repository.

For training, we generate a data set large enough for the models to converge, containing a total of $430\,080\,000$ reads. We follow an iterative procedure similar to the one described in \cite{rojas2019genet} to create mini-batches of reads. For each read in the mini-batch, its source genome is first selected uniformly at random. Then, the genome is used as input to the software \textit{InSilicoSeq}~\cite{insilicoseq} in order to generate the read. We simulate 151~bp error-free reads (default length of \textit{InSilicoSeq}) used for training, while we generate more realistic reads with Illumina NovaSeq-type noise for validation and evaluation of our models.
This approach allows us to evaluate the performance and generalization ability of our models in more realistic settings where sequencing noise is present. The evaluation of all methods was done on a total of $5\,242\,960$ NovaSeq type reads.

\begin{figure}
  \centering
  \includegraphics[width=1.0\linewidth]{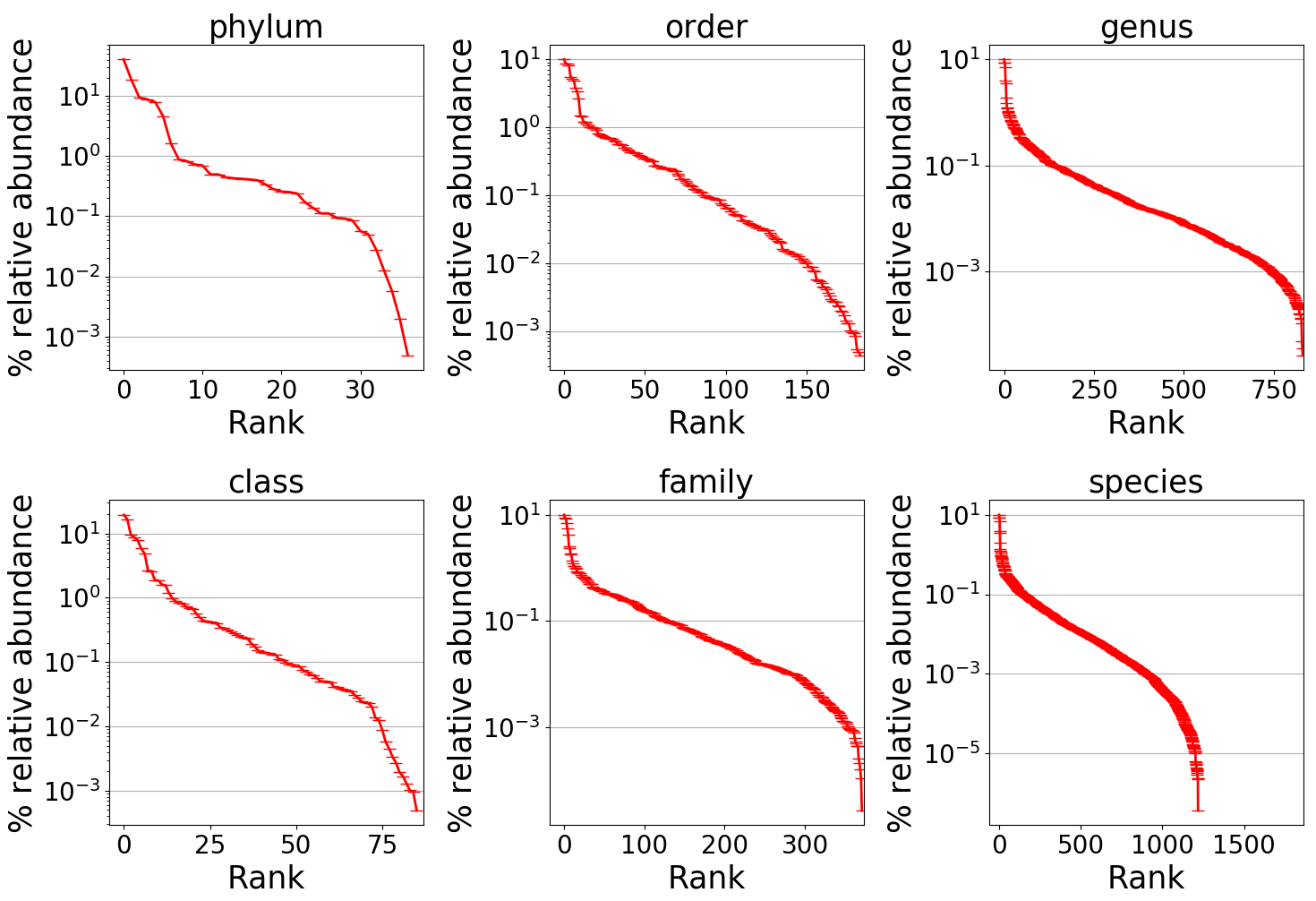}
  \caption{Rank-abundance curve for each taxonomic rank. All taxa are sorted using their abundance. Their abundance level is shown on the $y$-axis.}
  \label{fig:rankabundance}
\end{figure}

Training of the MIL models for abundance estimation is different, where a batch consists of a small number of bags of reads, each containing reads sampled using a more realistic distribution over the genomes. The procedure used is similar to the one used by the CAMISIM simulator~\cite{CAMISIM} and described in more detail in Section~\ref{sec:sampling_reads}. An example rank-abundance curve for each taxonomic rank generated by this procedure is shown in Figure~\ref{fig:rankabundance}.

Each bag is supposed to simulate a different microbial community, and hence, the generation procedure is repeated for each bag. The more realistic bags allow the MIL models to capture the interactions between the reads coming from related species and capture potential overlap in the reads originating from the same taxa.

Hyperparameter search was performed for all models. Details on the exact parameters can be found in Appendix~\ref{app:hyper}.

\subsubsection{Sampling a realistic set of reads}
\label{sec:sampling_reads}
In order to sample bags with a more realistic community of bacteria, we propose a method similar to the one by \cite{CAMISIM}. Given a set of all the taxa $\mathcal{T}$ at a higher level (e.g., \textit{genus} or \textit{family}, we sample $\left|\mathcal{T}\right|$ numbers from a lognormal distribution with $\mu=1$ and $\sigma=2$:

\begin{equation}
\label{lognormal}
    T_i \sim \mathop{Lognormal}(x;\mu, \sigma) = \frac{1}{x\sigma \sqrt{2\pi}} \exp{\left(-\frac{(\ln x - \mu)^2}{2\sigma^2}\right)}
\end{equation}

Then, for a taxon $t_i$ with $n$ genomes associated with it, we choose to include in our microbial community only $l_i$  random genomes where $l_i$ is sampled from a geometric distribution with $\mu=5$:

\begin{equation}
    P(X=l_i) = \left(1 - \frac{1}{\mu}\right)^{l_i} \frac{1}{\mu}
\end{equation}

To calculate the abundance of a genome $g_j$ belonging to taxon $t_i$, $l_i$ random numbers $Y_1 \dots Y_{l_i}$ are sampled from a lognormal distribution as in equation~(\ref{lognormal}). The abundance for the genome is then calculated as:
\begin{equation}
    A_j = \frac{Y_j}{\sum_{k=1}^{l_i} Y_k} T_i
\end{equation}

Finally, all abundances are normalized to produce a probability vector over all the genomes in the data set. When sampling a read set, a genome is selected by sampling from the distribution produced. Reads are then simulated from the genome sample using the software package \textit{InSilicoSeq}.

\subsection{Mapping-based baselines}
\subsubsection{Kraken 2}
\textit{Kraken 2}~\cite{wood2014kraken, wood2019improved} is a state-of-the-art taxonomic classification method that attempts to exactly match $k$-mers to reference genomes in its database. Kraken 2.0.8-beta was used for this study. \textit{Kraken 2} constructs a database containing pairs of the $l$-mer canonical minimizers of $k$-mers and their respective lowest common ancestors. Additionally, \textit{Kraken 2} allows for the user to specify a maximum hash table size in order to create databases that require less memory. This is achieved by a subsampling approach, reducing the number of minimizers included in the hash table. However, the accuracy of this method deteriorates significantly as the size of the database decreases. In addition, with larger databases, the overhead of loading the database into memory can be significant as shown in Table~\ref{tab:requirements}. 

The size of the complete database for our selected genomes is 12~GB. For comparison purposes, we also build \textit{Kraken} databases of three sizes in addition to the complete one: 8~GB, 500~MB, and 200~MB. The two smallest sizes in particular are comparable to the sizes of our proposed models.

For abundance estimation, we combine \textit{Kraken} with \textit{Bracken}~\cite{lu2017bracken}. \textit{Bracken} takes the classification results of \textit{Kraken} and re-estimates the frequencies of fragments assigned at lower taxonomic ranks by applying Bayes' theorem. \textit{Bracken} is especially useful in cases where a classification at a lower taxonomic rank (e.g., species) was not accurately determined by \textit{Kraken} but an assignment at a higher level (e.g., genus), was nevertheless determined. Bracken can heuristically assign reads in these cases to lower taxonomic ranks.

\subsubsection{Centrifuge}
\textit{Centrifuge}~\cite{centrifuge} compresses the genomes of all species by iteratively replacing pairs of most similar genomes with a much smaller compressed genome that excludes the identical parts shared by the original pair. Subsequently, an FM-index~\cite{fmindex} is built on top of the final genomic sequence. Classification is done by scoring exact matches of variable-length segments from the query sequences. For our selected reference genomes, the final database was 4.9 GB. Due to the nature of the compression method used, \textit{Centrifuge} does not provide any additional option of further restricting the database size. \textit{Centrifuge} version 1.0.4 was used in this study.

For abundance estimation, \textit{Centrifuge} directly produces a \textit{Kraken} style report containing the percentages for each taxon at various taxonomic ranks.

\subsection{Deep learning baselines}
\subsubsection{Genet}
\textit{GeNet} leverages the hierarchical nature of the taxonomy of species to simultaneously classify DNA reads at all taxonomic ranks~\cite{rojas2019genet}. The procedure is similar to positional embedding as described by~\cite{gehring2017convolutional}. Given an input $\vx = (x_1, \dots, x_n)$, an embedding $\vw=(\vw_1, \dots, \vw_n)$ is computed, where $\vw_i \in \sR^5$. The vocabulary of size $5$ corresponds to the symbols for the four possible nucleotides A, C, T, G, and N (for unknown base pairs in the read). Embeddings of the absolute positions for each letter are also computed to create $\vp=(\vp_1, \dots, \vp_n)$, where $\vp_i \in \sR^5$. The one-hot representation of the sequence, $\vo$, is added to the other two embeddings to create the matrix $\vw + \vp + \vo$. Subsequently, the resulting matrix is passed to a ResNet-like neural network which produces a final low-dimensional representation of the read. The main novelty of the architecture is the final layer used for classification, which comprises multiple softmax layers, one for each taxonomic rank. These layers are connected to each other so that information from higher ranks can be propagated towards the lower ranks. More formally, the output of softmax layer $i$ can be written as follows:
\begin{equation}
\vy_i = ReLU(\mW_i \vh) + ReLU(\mU_i \vy_{i-1}) \; ,
\end{equation}
where $\mW_i$ and $\mU_i$ are trainable parameters, $\vh_i$ is the output of the ResNet network and $\vy_{i-1}$ is the previous softmax output. $ReLU(\cdot)$ is the rectified linear unit function. To train the model, an averaged cross-entropy loss for each softmax layer is used.

\subsubsection{EmbedPool}
\cite{deepmicrobes} introduce multiple architectures for performing single-read classification among which the best is \textit{DeepMicrobes}. It involves embedding $k$-mers into a latent representation, followed by a bidirectional LSTM, a self-attention layer, and a multi-layer perceptron (MLP). Unlike \textit{GeNet}, this model can only be trained to classify a single taxonomic rank. Due to the significant amount of GPU memory required by the model, we implemented \textit{EmbedPool}, a simpler version of \textit{DeepMicrobes} (also described in the original paper) to use as a baseline. Because of the exclusion of an LSTM and a self-attention layer, \textit{EmbedPool} is faster and requires less GPU memory while still achieving similar accuracy~\cite{deepmicrobes}. 

\textit{EmbedPool} is a model that features an embedding layer for $k$-mers. In order to further reduce memory footprint, the embedding matrix only includes an entry for each canonical $k$-mer, approximately halving the memory requirements. Non-canonical $k$-mers use the embedding of their reverse complements. After the embedding is applied to each $k$-mer in a sequence, both max- and mean-pooling are performed on the resulting matrix and concatenated together to yield a low-dimensional representation of the read. Since the embedding dimension is set to $100$, after concatenation, this results in a vector of size $200$. An MLP with one hidden layer of $3\,000$ units subsequently classifies the read. ReLU is used as the activation function. The model is trained end-to-end using cross-entropy loss.

In order to classify at multiple taxonomic ranks, one could run multiple instances of the model, each running on a different GPU. Even though \textit{EmbedPool} has a lower GPU memory footprint, the requirements can still scale very quickly. When setting $k=12$, the model requires 3.26~GB of memory. Just increasing $k$ to $k=13$ or $k=14$, the model size significantly increases up to 13~GB  and 50~GB. The training requirements can even be much higher than that due to memory required during backpropagation or for storing intermediate values, making these models very difficult to train~\cite{shoeybi2019megatron} or to fit in GPU memory.

\subsection{Proposed memory-efficient models}
In this section, we describe two novel models based on \textit{EmbedPool} that are very memory-efficient while achieving comparable performance when compared to both deep-learning and mapping-based baselines.

\begin{figure}
\subfigure[LSH-EmbedPool]{
    \includegraphics[scale=0.65]{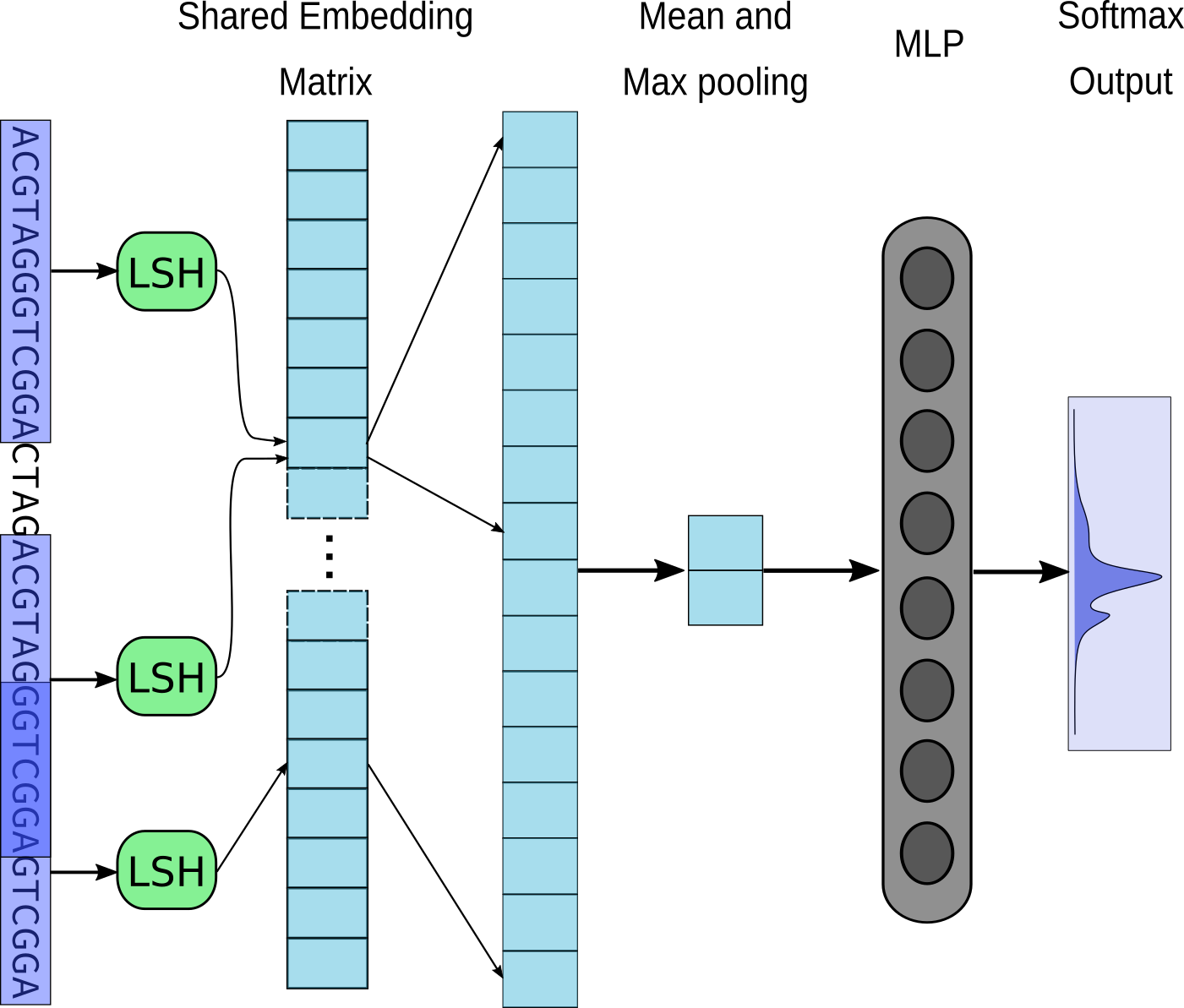}
    \label{fig:lshembedpool}
}\hfill
\subfigure[Hash-EmbedPool]{
    \includegraphics[scale=0.65]{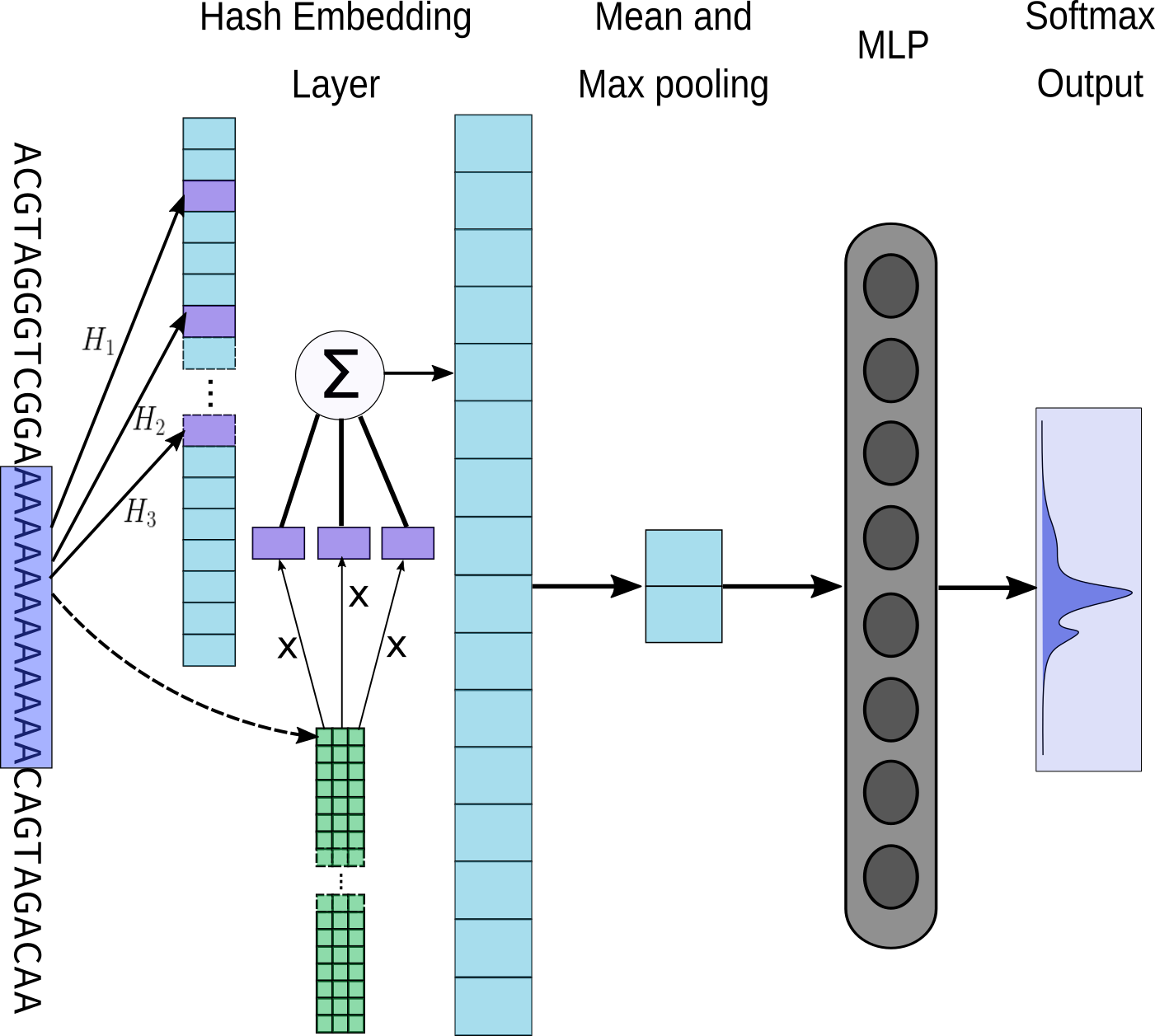}
    \label{fig:hashembedpool}
}
\caption{The two proposed memory-efficient architectures for single read classification. \textit{LSH-EmbedPool} uses a locality-sensitive hash to place each $k$-mer into buckets. In this example, two of the $k$-mers are similar, with only one nucleotide being different. Therefore, the same embedding is used to represent both. This achieves significant memory reductions since multiple $k$-mers can be represented by the same embedding. \textit{Hash-EmbedPool} uses multiple hash functions to select embeddings from a shared embedding pool where each selected vector is weighted by trainable weights. The weighted sum is then used as the final representation of the $k$-mer.}
\label{fig:singlearchs}
\end{figure}

\subsubsection{LSH-EmbedPool}
As explained, the largest portion of the memory is used for storing the embedding matrix, which can make the models unusable for larger values of $k$. However, a large $k$ can have a significant effect on the performance of the resulting model. A simple, but less effective way to fix this problem is to first fix the vocabulary size to a predefined number $h\in \sZ$ such that $h < 4^k$. Then a hash function could be used to map each $k$-mer to an embedding that is shared by all $k$-mers that collide in the same bucket. This is also known as \textit{feature hashing} \cite{weinberger2009feature}. More formally, a hash function is a function $H: \Sigma^k \rightarrow \sZ_{h} $ where $\Sigma^k$ is the set of all $k$-mers and $\sZ_h$ is the set of integers modulo $h$. Given a $k$-mer $w$, then $H(w)$ can be used as an index to the embedding matrix. The problem with the described approach is that collisions can be detrimental to the training of the model since two very disimilar $k$-mers can potentially be mapped to the same embedding vector, each pushing the gradient during training towards completely different directions. Furthermore, in order to keep the size of the model small and since each embedding vector has a size of $100$, $h$ cannot be much larger than $2^{20}$ because this would again significantly increase the GPU requirements. This results in a high probability that all $k$-mers would collide with at least one or more other $k$-mers.

From our informal analysis above, it is clear that two approaches can be taken to reduce the burden of collisions: a) reduce the number of collisions, or b) avoid collision of dissimilar $k$-mers. \textit{LSH-EmbedPool} follows the second approach by replacing the hash function with a locality-sensitive hashing (LSH) scheme~\cite{gionis1999similarity,indyk1998approximate}. An LSH is a function that, given two very similar inputs, outputs the same hash with high probability. This is in contrast to the regular hash function where the collisions happen completely at random. LSHs have been widely used in near-duplicate detection for documents in natural language processing~\cite{szmit2013locality} as well as by biologists for finding similar gene expressions in genome-wide association studies~\cite{brinza2010rapid}. For our case, we use an LSH based on MinHash. In particular we use the software \textit{sourmash}~\cite{brown2016sourmash, broder1997resemblance} where as a first step a list of hashes of all $l$-mers in a $k$-mer (where $l < k$) is created. Subsequently, the resulting hashes are sorted numerically in ascending order and the second half of the list is discarded. The remaining sorted hashes are combined by one more application of a fast non-cryptographic hash function. For this purpose, the library \textit{xxHash} is used~\cite{xxhash}. Discarding some of the hashes allows the algorithm to tolerate minor differences in $k$-mers, such as single nucleotide changes, and increases the probability of such $k$-mers being hashed to the same bucket. Furthermore, the parameter $l$ can be used to control the sensitivity of the scheme, with smaller values resulting in a more sensitive scheme. In our experiments, we found that setting $l$ to be approximately half of $k$ is ideal, while our best accuracy was achieved by setting $k=13$ and $l=7$. In order to keep memory requirements to the minimum, we set the vocabulary size $h=2^{20}$, which results in a model of only 457.96~MB. This is a reduction of a factor of $29$ compared to using a standard \textit{EmbedPool} with $k=13$. As with the standard \textit{EmbedPool}, the model is trained end-to-end using cross-entropy loss. Figure~\ref{fig:lshembedpool} shows an overview of our architecture.

\subsubsection{Hash-EmbedPool}
\label{sec:hashembed}
In contrast to \textit{LSH-EmbedPool}, \textit{Hash-EmbedPool} follows the approach of reducing collisions. This can be achieved using hash embeddings, first introduced by \cite{svenstrup2017hash}. Hash embeddings are a hybrid between a standard embedding and the naive feature hashing method. As with \textit{LSH-EmbedPool}, there is a set of $h$ shared embedding vectors. For each $k$-mer, we use $q$ different hash functions $H_0,\dots,H_{q-1} : \Sigma^k \rightarrow \sZ_h$, to select $q$ embedding vectors from the pool of shared embeddings. The final vector representation of the $k$-mer is a weighted sum of the selected vectors using $q$ weights $p_1,\dots, p_q$. The weights for each possible $k$-mer are stored in a second learnable matrix $\mW$ of size $4^k \cdot q$. More formally, let $D: \Sigma^k \rightarrow \sZ_{4^k}$ be a function that maps each $k$-mer to its unique index and $E: \sZ_h \rightarrow \sR^d$ be the function that returns shared embedding vectors given an integer in $\sZ_h$. The final embedding of a $k$-mer $w$ can be written as $e_w=\sum_{i=0}^{q} p_i E \left(H_i(w) \right)$ where $p_i = \mW_{D(w),i}$. Intuitively, even though collisions can occur in the shared embedding matrix, because of the unique weighted sum of the vectors for each different $k$-mer, collisions in the final embedding are reduced significantly. Increasing $q$ or $h$ of course has the effect of further reducing the number of collisions at the cost of higher memory requirements. In our model, we found $q=3$ to be sufficient while $h$ was set to $2^{20}$ to keep memory requirements low. We found $k=12$ to be optimal for this model, resulting in a final model size of 553.90~MB, a reduction of a factor of $6$ from a standard \textit{EmbedPool} using the same $k$ parameter. For the hash functions $H_0,\dots,H_{q-1}$ we randomly sample functions from the hash family proposed by \cite{carter1979universal}. As with the standard \textit{EmbedPool}, the model is trained end-to-end using cross-entropy loss. An overview of the architecture is shown in Figure~\ref{fig:hashembedpool}.

\subsection{Proposed MIL methods}

In this section, we present new models for abundance estimation in metagenomic read sets.
Since the co-occurence of species can be very informative in real-world samples (see Fig.~\ref{fig:rankabundance}), we combine the presented deep learning approaches with permutation-invariant aggregators for multiple instance learning.

\begin{figure*}[htp]

\subfigure[GeNet + MIL pooling]{
    \includegraphics[scale=0.75]{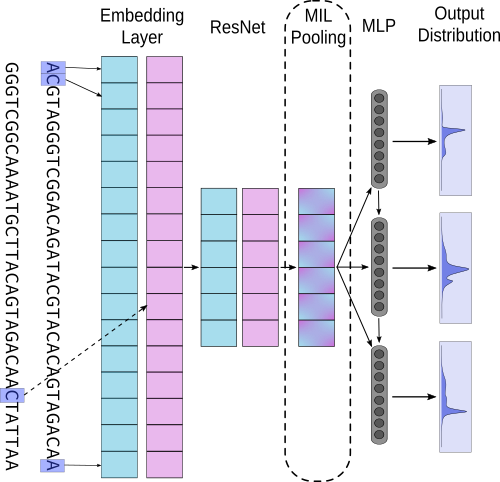}
    \label{fig:archsa}
}\hfill
\subfigure[EmbedPool + MIL pooling]{
    \includegraphics[scale=0.75]{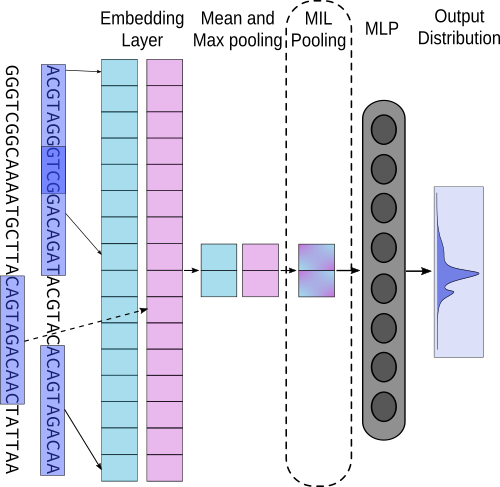}
    \label{fig:archsb}
}

\caption{The two proposed architectures for solving the MIL task. The models can process multiple reads (only two reads shown for compactness) independently from each other. During the \textit{MIL pooling} phase, the outputs for each read are combined to create a representation for the whole read set. Subsequently, the model can use this to directly predict the distribution over the taxa.}
\label{fig:archs}
\end{figure*}

\subsubsection{GeNet + MIL pooling}
\label{sec:GenetMIL}
A mini-batch of bags of reads is used as input. The first part of \textit{GeNet}, consisting of the embedding and the ResNet, is used to process each read individually. A pooling layer is then used to group all reads in each bag to create bag-level embeddings. This is also referred to as MIL pooling~\cite{MILDefinition, MILSurvey}. The output is passed to the final layers of \textit{GeNet} in order to output a probability distribution over the taxa at each taxonomic rank. As a loss function we use the Jensen-Shannon ($D_{JS}$) divergence~\cite{JSdivergence} between the predicted distribution and the actual distribution of the bag.

Given that a bag is a set, we require that a MIL pooling layer is permutation-invariant, that is, permuting the reads of the bag should still produce the same result. To this end, we utilize DeepSets~\cite{deepsets}. DeepSets can be formally described as follows:
\begin{equation}
    f(X) = \rho \left( \sum_{x \in X}\phi(x) \right) \;.
\end{equation}
In other words, each element of a set $X$ is first processed by a function $\phi(\cdot)$. The outputs are all summed together and the result is subsequently transformed by a function $\rho(\cdot)$. \cite{deepsets} proved that all valid functions operating on subsets of countable sets or on fixed-sized subsets of uncountable sets can be written in this form. In our case, the inputs are embeddings in $\sR^{5 \times L}$ where $L$ is the length of a read. In addition, we only input bags of fixed size and hence the assumptions of Theorem 2 in \cite{deepsets} are satisfied. $\rho(\cdot)$ is modelled with a small MLP with one hidden layer while the ResNet part of the network models the function $\phi(\cdot)$. 

Alternatively to DeepSets, we also consider an attention-based pooling layer as seen in \cite{attentionMIL}, motivated by the fact that it would allow the model to attend to specific reads originating from each species. In attention-based pooling, the elements of the set are combined in different ways to create a set $\vz = \{\vz_1,\dots, \vz_k\}$, such that the set remains invariant when we permute the elements of the input set. This can be written as follows:
\begin{equation}
    \vz_j = \sum_{k=1}^K \alpha_{j,k} \vx_k \; ,
\end{equation}

\begin{equation}
    \alpha_{j,k} = \frac{\exp( \vw_j^T \tanh(\mV \vx_k^T) )}{\sum_{l=1}^K \exp( \vw_j^T \tanh(\mV \vx_l^T) )} \; ,
\end{equation}
where $\vx_k$ is an element of the input set, and $\mV$ and $\vw_j$ are trainable parameters. The weights $\alpha_{j,k}$ are therefore calculated with an MLP with one hidden layer with $\tanh$ non-linearity and $\softmax$ activation at the end. \cite{attentionMIL} also attempt to increase the flexibility of the MIL pooling by introducing a gating mechanism as shown below:
\begin{equation}
    \alpha_{j,k} = \frac{\exp( \vw_j^T \left( \tanh(\mV \vx_k^T) \odot \sigmoid(\mU \vx_k^T) \right) )}{\sum_{l=1}^K \exp( \vw_j^T \left( \tanh(\mV \vx_l^T) \odot \sigmoid(\mU \vx_k^T) \right) )} \; ,
\end{equation}
where $\mU$ is an additional learnable matrix, $\sigmoid$ is the sigmoid activation function and $\odot$ is the element-wise product. As shown in Appendix~\ref{app:hyper}, for our models, using the gating mechanism is an additional hyperparameter. Following the attention mechanism, the output $\vz$ is flattened to create a single vector for each bag which is subsequently processed by \textit{GeNet}'s final layers to output the predicted distributions. The overall architecture can be seen in Figure~\ref{fig:archsa}.

\subsubsection{EmbedPool + MIL pooling}
Similarly to subsection \ref{sec:GenetMIL}, we use \textit{EmbedPool} to process the reads individually. A MIL pooling layer is added after the mean- and max- pooling layers, the output of which is fed to the rest of the model to predict the distribution. JS-divergence is used as a loss function. For MIL pooling, we use DeepSets and attention-based pooling as before. An overview of the model can again be seen in Figure~\ref{fig:archsb}.

\subsubsection{Hash-EmbedPool + MIL pooling}
We follow the same approach as \textit{EmbedPool + MIL pooling}, but each read is instead processed by a Hash-EmbedPool as introduced in section~\ref{sec:hashembed}.

\section{Results and Discussion}
\label{sec:results}

\subsection{Single read classification}
In this section, we analyze the memory requirements and performance of our proposed memory-efficient models for taxonomic classification and compare them with our baselines. The evaluation of all models was done on a total of $5\,242\,960$ reads with Illumina NovaSeq-type noise.

Our best performing model is \textit{LSH-EmbedPool} with an accuracy of $0.739$, outperforming both \textit{GeNet} and standard \textit{EmbedPool} which scored $0.111$ and $0.431$, respectively. In \cite{rojas2019genet}, \textit{GeNet} was trained on PacBio reads of length $10\,000$~bp and Illumina reads of length $1\,000$~bp. Since in most cases genome sequencing technologies like Illumina produce shorter reads in the range of 100~bp~-~300~bp~\cite{quail2012tale}, we chose to train all our models on reads of length 151~bp. This difference explains the discrepancy between our results and the results reported by \cite{rojas2019genet}. Since \textit{GeNet} uses one-hot encoding rather than $k$-mer encoding, it might be unable to extract useful features shared across the whole genome from shorter reads. Similarly, \textit{Hash-EmbedPool} outperforms both deep learning baselines. Compared to \textit{Kraken}, our models achieve significantly higher accuracy when \textit{Kraken}'s database is subsampled to have similar size to our models. At 500~MB, \textit{Kraken} only achieves an accuracy of $0.646$. Reducing the database even further, \textit{Kraken}'s accuracy is $0.371$, even lower than our deep learning baseline \textit{EmbedPool}. Note that \textit{EmbedPool} was trained with $k=8$, only requiring a memory of 70.51~MB. Furthermore, the top-5 accuracy achieved by \textit{LSH-EmbedPool} is comparable to the accuracy of \textit{Kraken} restricted to 8~GB. \textit{Centrifuge} achieves the highest classification performance, with a database size of 4.9~GB. However, this comes at the cost of a reduction in speed, with \textit{Centrifuge} only being able to process $278\,000$ reads per minute when run on 4 threads. Comparably, \textit{Hash-EmbedPool} can process $342\,000$ reads/min when the time required to pre-process reads is measured, and $607\,000$ reads/min when that time is excluded, comparable to \textit{Kraken}. However, \textit{Kraken}'s simple exact $k$-mer matching technique makes it the fastest even when the time required to load the whole database into memory is taken into account. As \textit{Kraken}'s database size is reduced, the classification speed increases and the time required to load the database is less significant. Note that for the deep learning models, a single GPU was used. Classification speed can scale linearly when more GPUs are used to process multiple mini-batches of reads simultaneously. Our results demonstrate that the proposed models effectively learn memory-efficient representations while demonstrating a tradeoff between classification accuracy, speed, and memory. Classification accuracy of all models is summarized in Table~\ref{tab:accuracy}, while speed and memory requirements are shown in Table~\ref{tab:requirements}

\begin{table}[t!]
\centering
\caption{Accuracy comparison for single read species classification between different machine learning and conventional methods. When compared under the same memory restrictions (model size $<$ 600~MB), our proposed methods outperform the baselines. Methods outside of that memory restriction are shown in the lower part of the table for comparison purposes.}
\label{tab:accuracy}
\begin{tabular}{@{}lcc@{}}
\toprule
 & Accuracy & Top 5 accuracy \\ \midrule
GeNet & $0.111 \pm 0.016$ & $0.229 \pm 0.008$ \\
EmbedPool & $0.431 \pm 0.010$ & $0.612 \pm 0.009$ \\
Kraken 200 MB & $0.371 \pm 0.006$ & - \\
Kraken 500 MB & $0.646 \pm 0.005$ & - \\
Hash-EmbedPool (ours) & $0.719 \pm 0.003$ & $0.797 \pm 0.003$ \\
LSH-EmbedPool (ours) & $\mathbf{0.739 \pm 0.003}$ & $\mathbf{0.814 \pm 0.003}$ \\
\midrule
Kraken 8 GB & $0.821 \pm 0.005$ & - \\
Kraken & $0.897 \pm 0.004$ & - \\
Centrifuge & $0.929 \pm 0.003$ & - \\ \bottomrule
\end{tabular}
\end{table}

\begin{table}[t!]
\centering
\caption{Comparison of runtimes and model sizes for the different methods on single read classification. Kraken is faster than the other methods, but this difference decreases with the size of the model. As seen in Table~\ref{tab:accuracy}, the small versions of Kraken actually perform very poorly. Centrifuge performs well (see Tab.~\ref{tab:accuracy}), but runs more slowly and has a larger model size. Our proposed methods trade off a slightly lower accuracy for major improvements in model size and speed.}
\label{tab:requirements}
\begin{tabular}{@{}lccc@{}}
\toprule
 & Parameters & Model Size & Kseqs/min \\ \midrule
GeNet & $16\,110\,528$ & 61.46~MB & 561 - 577 \\
EmbedPool & $18\,483\,562$ & 70.51~MB & 342 - 607 \\ 
Hash-EmbedPool (ours) & $145\,223\,433$ & 553.90~MB & 342 - 607 \\
LSH-EmbedPool (ours) & $120\,051\,462$ & 457.96~MB & 147 - 607 \\
\midrule
Kraken 200 MB & - & 200~MB & 31,418 - 43,048 \\
Kraken 500 MB & - & 500~MB & 23,077 - 39,491 \\
Kraken 8 GB & - & 8~GB &  4,769 - 23,441 \\
Kraken & - & 12~GB & 653 - 13,275 \\
Centrifuge & - & 4.9~GB & 275 - 278 \\ \bottomrule
\end{tabular}
\end{table}

\subsection{Abundance estimation with MIL pooling}

\begin{figure}[t!]
  \centering
  \includegraphics[width=0.8\linewidth]{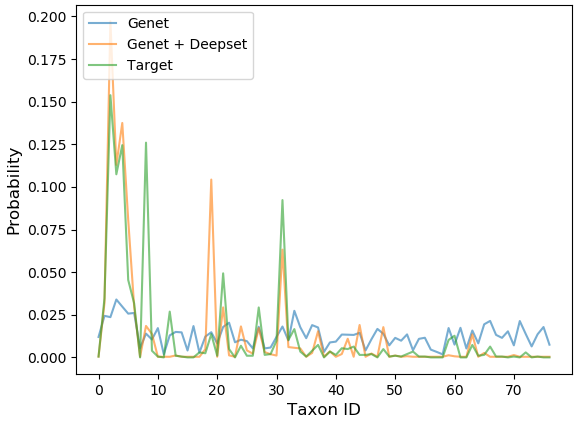}
  \caption{Distribution of taxa at the class rank. The target distribution is denoted in orange and the output of the model is denoted in blue.}
  \label{fig:milexample}
\end{figure}

Following the comparison of our methods for taxonomic classification, we proceed to the related problem of abundance estimation. For the standard \textit{GeNet} and \textit{EmbedPool}, the microbiota distribution was calculated by classifying each read independently, while for our proposed models, the distribution was predicted directly by the models. An example of the output of the MIL models is shown in Figure~\ref{fig:milexample}.

All models were evaluated on a total of $100$ bags of $1\,024\,000$ NovaSeq-type reads each. Both \textit{GeNet + Deepset} and \textit{GeNet + Attention} perform better than all other deep learning models and comparably to the mapping-based methods at higher taxonomic ranks. Moreover, this accuracy is achieved with a model size 4 times smaller compared to the smallest Kraken baseline and 200 times smaller compared to the largest one. Notably, Kraken's abundance estimation ability does not decrease significantly at higher taxonomic ranks when the database size is restricted. This is due to the fact that Kraken still has enough correctly classified reads to calculate an approximation of the true distribution after discarding all the unclassified reads. Similarly, we found that increasing the size of the deep learning models does not seem to always improve the abundance estimation performance, even though single-read accuracy increases significantly (as was also shown by~\cite{deepmicrobes}).

\begin{table*}[ht!]
\centering
\caption{JS-divergence for all models. Our MIL models achieve superior performance compared to other deep-learning models at higher taxonomic ranks up to \textit{Family} and significantly reduce the memory requirements compared to mapping-based methods.}
\label{tab:js}
\begin{tabular}{@{}lcccccc@{}}
\toprule
 & Phylum & Family & Species \\ \midrule
GeNet~\cite{rojas2019genet} & $0.412 \pm 0.057$ & $0.614 \pm 0.063$ & $0.920 \pm 0.059$ \\
EmbedPool~\cite{deepmicrobes} & $0.105 \pm 0.027$ & $0.457 \pm 0.081$ & $0.741 \pm 0.067$ \\ \midrule
GeNet + Deepset (ours) & $0.053 \pm 0.029$ & $0.416 \pm 0.131$ & $1.115 \pm 0.253$ \\
GeNet + Attention (ours) & $0.058 \pm 0.030$ & $0.446 \pm 0.145$ & $1.140 \pm 0.251$ \\
EmbedPool + Deepset (ours) & $0.065 \pm 0.041$ & $0.540 \pm 0.114$ & $1.101 \pm 0.227$ \\
EmbedPool + Attention (ours) & $0.065 \pm 0.041$ & $0.539 \pm 0.113$ & $1.105 \pm 0.245$ \\ 
Hash-EmbedPool + Deepset (ours) & $0.059 \pm 0.040$ & $0.537 \pm 0.112$ & $1.201 \pm 0.203$ \\ 
Hash-EmbedPool + Attention (ours) & $0.060 \pm 0.040$ & $0.538 \pm 0.111$ & $1.207 \pm 0.211$ \\ \midrule
Kraken 200~MB & $0.037 \pm 0.021$ & $0.373 \pm 0.054$ & $0.642 \pm 0.078$ \\
Kraken 500~MB & $0.015 \pm 0.008$ & $0.188 \pm 0.049$ & $0.377 \pm 0.061$ \\
Kraken 8~GB & $0.012 \pm 0.014$ & $0.129 \pm 0.038$ & $0.333 \pm 0.094$ \\
Kraken & $0.007 \pm 0.003$ & $0.098 \pm 0.013$ & $0.237 \pm 0.056$ \\
Centrifuge & $0.00007 \pm 0.000001$ & $0.0014 \pm 0.0005$ & $0.030 \pm 0.020$ \\ \bottomrule
\end{tabular}
\end{table*}

As explained in Section~\ref{sec:intro}, we believe that the improvement in performance at higher taxonomic ranks is owed to the fact that the models can exploit the co-occurrence of species in realistic settings or detect overlaps of reads in a bag. A drawback of our MIL models is that, since the performance is owed to the special structure of the bags, it is unlikely that they would perform well when presented with bags with an unrealistic distribution of species (e.g., a bag with a uniformly random distribution over all species). Therefore, it is clear that the models achieve a trade-off between flexibility and performance. Moreover, our proposed MIL models perform poorly on the finer taxonomic ranks, possibly because in the MIL setting the models only observe a summary of the bag rather than a label for each instance. It might therefore be harder for them to learn adequate features. However, the greater performance on higher levels can prove beneficial for some real-world metagenomic data sets where reference data is not sufficiently available to train deep learning models accurately~\cite{afshinnekoo2015geospatial,tully2018reconstruction}. Table~\ref{tab:js} shows a comparison of our MIL models and the achieved $D_{JS}$ values. A comparison of \textit{GeNet + Deepset} (our best performing model) and standard \textit{GeNet} can be seen in Figure~\ref{fig:genetvsds} in Appendix~\ref{app:supp_results}.

\section{Conclusions}
In this work, we introduced memory-efficient models for taxonomic classification. In particular, our method for combining locality-sensitive hashing with learnable embeddings is general and could be used for other tasks where the large size of the embedding matrix would otherwise hinder the training and usability of the model. Our proposed architectures outperformed all other methods within the same memory restrictions and were comparable to mapping-based methods that require larger amounts of memory.

Subsequently, we tackled the problem of directly predicting the distribution of the microbiota in metagenomic samples. In contrast to previous methods that are based on classifying single reads, we formulated the problem as a multiple instance learning task and used permutation-invariant pooling layers in order to learn low-dimensional embeddings for whole sets of reads. We showed that our proposed method can perform better than the baseline deep learning models and comparable to mapping-based methods at the higher taxonomic ranks. Moreover, our best performing MIL models were significantly smaller compared to the mapping-based methods. The MIL models presented could be used as an initial step to filter or preselect the potential genomes that more traditional alignment methods would need to take as input in order to increase their performance.

Future work could include exploring alternative base architectures or more sophisticated pooling methods that can better capture the interactions between reads. For example, one could use Janossy pooling~\cite{murphy2018janossy}, another permutation-invariant method that can capture $k^\text{th}$ order interactions between the elements of a set.
Also, the models could potentially be combined with a probabilistic component, such as a Gaussian process over DNA sequences~\cite{fortuin2018scalable}, to allow for uncertainty estimates on the predictions.
Finally, as previously mentioned, a possible issue is that observing only the summary of the read set can make it more difficult for the model to learn adequate features for the individual reads. A solution to this could be to first learn better instance-level embeddings to use as input, in order to aid the model in learning more suitable bag-level embeddings.

\bibliographystyle{plain}
\bibliography{references}

\begin{thebibliography}{10}

\bibitem{afshinnekoo2015geospatial}
Ebrahim Afshinnekoo, Cem Meydan, Shanin Chowdhury, Dyala Jaroudi, Collin Boyer,
  Nick Bernstein, Julia~M Maritz, Darryl Reeves, Jorge Gandara, Sagar
  Chhangawala, et~al.
\newblock Geospatial resolution of human and bacterial diversity with
  city-scale metagenomics.
\newblock {\em Cell systems}, 1(1):72--87, 2015.

\bibitem{almodaresi2019efficient}
Fatemeh Almodaresi, Prashant Pandey, Michael Ferdman, Rob Johnson, and Rob
  Patro.
\newblock An efficient, scalable and exact representation of high-dimensional
  color information enabled via de bruijn graph search.
\newblock In {\em International Conference on Research in Computational
  Molecular Biology}, pp. 1--18. Springer, 2019.

\bibitem{bingmann2019cobs}
Timo Bingmann, Phelim Bradley, Florian Gauger, and Zamin Iqbal.
\newblock Cobs: a compact bit-sliced signature index.
\newblock In {\em International Symposium on String Processing and Information
  Retrieval}, pp. 285--303. Springer, 2019.

\bibitem{bradley2019ultrafast}
Phelim Bradley, Henk~C den Bakker, Eduardo~PC Rocha, Gil McVean, and Zamin
  Iqbal.
\newblock Ultrafast search of all deposited bacterial and viral genomic data.
\newblock {\em Nature biotechnology}, 37(2):152, 2019.

\bibitem{brady2009phymm}
Arthur Brady and Steven~L Salzberg.
\newblock Phymm and phymmbl: metagenomic phylogenetic classification with
  interpolated markov models.
\newblock {\em Nature methods}, 6(9):673, 2009.

\bibitem{brinza2010rapid}
Dumitru Brinza, Matthew Schultz, Glenn Tesler, and Vineet Bafna.
\newblock Rapid detection of gene--gene interactions in genome-wide association
  studies.
\newblock {\em Bioinformatics}, 26(22):2856--2862, 2010.

\bibitem{broder1997resemblance}
Andrei~Z Broder.
\newblock On the resemblance and containment of documents.
\newblock In {\em Proceedings. Compression and Complexity of SEQUENCES 1997
  (Cat. No. 97TB100171)}, pp. 21--29. IEEE, 1997.

\bibitem{brown2016sourmash}
C~Brown and Luiz Irber.
\newblock sourmash: a library for minhash sketching of dna.
\newblock {\em Journal of Open Source Software}, 1(5):27, 2016.

\bibitem{busia2019deep}
Akosua Busia, George~E Dahl, Clara Fannjiang, David~H Alexander, Elizabeth
  Dorfman, Ryan Poplin, Cory~Y McLean, Pi-Chuan Chang, and Mark DePristo.
\newblock A deep learning approach to pattern recognition for short dna
  sequences.
\newblock {\em bioRxiv}, p. 353474, 2019.

\bibitem{MILSurvey}
Marc-Andr{\'e} Carbonneau, Veronika Cheplygina, Eric Granger, and Ghyslain
  Gagnon.
\newblock Multiple instance learning: A survey of problem characteristics and
  applications.
\newblock {\em Pattern Recognition}, 77:329--353, 2018.

\bibitem{carter1979universal}
J~Lawrence Carter and Mark~N Wegman.
\newblock Universal classes of hash functions.
\newblock {\em Journal of computer and system sciences}, 18(2):143--154, 1979.

\bibitem{xxhash}
Yann Collet.
\newblock {xxHash - Extremely fast hash algorithm}.
\newblock \url{https://github.com/Cyan4973/xxHash}, 2012.
\newblock [Online; accessed 30-January-2020].

\bibitem{metasub2016metagenomics}
MetaSUB~International Consortium et~al.
\newblock The metagenomics and metadesign of the subways and urban biomes
  (metasub) international consortium inaugural meeting report, 2016.

\bibitem{danko2019global}
David Danko, Daniela Bezdan, Ebrahim Afshinnekoo, Sofia Ahsanuddin, Chandrima
  Bhattacharya, Daniel~J Butler, Kern~Rei Chng, Francesca De~Filippis, Jochen
  Hecht, Andre Kahles, et~al.
\newblock Global genetic cartography of urban metagenomes and anti-microbial
  resistance.
\newblock {\em BioRxiv}, p. 724526, 2019.

\bibitem{fmindex}
Paolo Ferragina and Giovanni Manzini.
\newblock Opportunistic data structures with applications.
\newblock In {\em Proceedings 41st Annual Symposium on Foundations of Computer
  Science}, pp. 390--398. IEEE, 2000.

\bibitem{fortuin2018scalable}
Vincent Fortuin, Gideon Dresdner, Heiko Strathmann, and Gunnar R{\"a}tsch.
\newblock Scalable gaussian processes on discrete domains.
\newblock {\em arXiv preprint arXiv:1810.10368}, 2018.

\bibitem{MILDefinition}
James Foulds and Eibe Frank.
\newblock A review of multi-instance learning assumptions.
\newblock {\em The Knowledge Engineering Review}, 25(1):1--25, 2010.

\bibitem{CAMISIM}
Adrian Fritz, Peter Hofmann, Stephan Majda, Eik Dahms, Johannes Dr{\"o}ge,
  Jessika Fiedler, Till~R Lesker, Peter Belmann, Matthew~Z DeMaere, Aaron~E
  Darling, et~al.
\newblock {CAMISIM}: simulating metagenomes and microbial communities.
\newblock {\em Microbiome}, 7(1):17, 2019.

\bibitem{ganscha2018supervised}
Stefan Ganscha, Vincent Fortuin, Max Horn, Eirini Arvaniti, and Manfred
  Claassen.
\newblock Supervised learning on synthetic data for reverse engineering gene
  regulatory networks from experimental time-series.
\newblock {\em bioRxiv}, p. 356477, 2018.

\bibitem{garrison2018variation}
Erik Garrison, Jouni Sir{\'e}n, Adam~M Novak, Glenn Hickey, Jordan~M Eizenga,
  Eric~T Dawson, William Jones, Shilpa Garg, Charles Markello, Michael~F Lin,
  et~al.
\newblock Variation graph toolkit improves read mapping by representing genetic
  variation in the reference.
\newblock {\em Nature biotechnology}, 36(9):875--879, 2018.

\bibitem{gehring2017convolutional}
Jonas Gehring, Michael Auli, David Grangier, Denis Yarats, and Yann~N Dauphin.
\newblock Convolutional sequence to sequence learning.
\newblock In {\em Proceedings of the 34th International Conference on Machine
  Learning-Volume 70}, pp. 1243--1252. JMLR. org, 2017.

\bibitem{gionis1999similarity}
Aristides Gionis, Piotr Indyk, Rajeev Motwani, et~al.
\newblock Similarity search in high dimensions via hashing.
\newblock In {\em Vldb}, volume~99, pp. 518--529, 1999.

\bibitem{insilicoseq}
Hadrien Gourl{\'e}, Oskar Karlsson-Lindsj{\"o}, Juliette Hayer, and Erik
  Bongcam-Rudloff.
\newblock Simulating illumina metagenomic data with insilicoseq.
\newblock {\em Bioinformatics}, 35(3):521--522, 2018.

\bibitem{howe2014tackling}
Adina~Chuang Howe, Janet~K Jansson, Stephanie~A Malfatti, Susannah~G Tringe,
  James~M Tiedje, and C~Titus Brown.
\newblock Tackling soil diversity with the assembly of large, complex
  metagenomes.
\newblock {\em Proceedings of the National Academy of Sciences},
  111(13):4904--4909, 2014.

\bibitem{huson2007megan}
Daniel~H Huson, Alexander~F Auch, Ji~Qi, and Stephan~C Schuster.
\newblock Megan analysis of metagenomic data.
\newblock {\em Genome research}, 17(3):377--386, 2007.

\bibitem{attentionMIL}
Maximilian Ilse, Jakub~M Tomczak, and Max Welling.
\newblock Attention-based deep multiple instance learning.
\newblock {\em arXiv preprint arXiv:1802.04712}, 2018.

\bibitem{indyk1998approximate}
Piotr Indyk and Rajeev Motwani.
\newblock Approximate nearest neighbors: towards removing the curse of
  dimensionality.
\newblock In {\em Proceedings of the thirtieth annual ACM symposium on Theory
  of computing}, pp. 604--613, 1998.

\bibitem{karasikov2019sparse}
Mikhail Karasikov, Harun Mustafa, Amir Joudaki, Sara Javadzadeh-No, Gunnar
  R{\"a}tsch, and Andr{\'e} Kahles.
\newblock Sparse binary relation representations for genome graph annotation.
\newblock In {\em International Conference on Research in Computational
  Molecular Biology}, pp. 120--135. Springer, 2019.

\bibitem{centrifuge}
Daehwan Kim, Li~Song, Florian~P Breitwieser, and Steven~L Salzberg.
\newblock Centrifuge: rapid and sensitive classification of metagenomic
  sequences.
\newblock {\em Genome research}, 26(12):1721--1729, 2016.

\bibitem{la2015probabilistic}
Massimo La~Rosa, Antonino Fiannaca, Riccardo Rizzo, and Alfonso Urso.
\newblock Probabilistic topic modeling for the analysis and classification of
  genomic sequences.
\newblock {\em BMC bioinformatics}, 16(6):S2, 2015.

\bibitem{deepmicrobes}
Qiaoxing Liang, Paul~W Bible, Yu~Liu, Bin Zou, and Lai Wei.
\newblock Deepmicrobes: taxonomic classification for metagenomics with deep
  learning.
\newblock {\em bioRxiv}, p. 694851, 2019.

\bibitem{JSdivergence}
Jianhua Lin.
\newblock Divergence measures based on the shannon entropy.
\newblock {\em IEEE Transactions on Information theory}, 37(1):145--151, 1991.

\bibitem{lu2017bracken}
Jennifer Lu, Florian~P Breitwieser, Peter Thielen, and Steven~L Salzberg.
\newblock Bracken: estimating species abundance in metagenomics data.
\newblock {\em PeerJ Computer Science}, 3:e104, 2017.

\bibitem{luo2019metagenomic}
Yunan Luo, Yun~William Yu, Jianyang Zeng, Bonnie Berger, and Jian Peng.
\newblock Metagenomic binning through low-density hashing.
\newblock {\em Bioinformatics}, 35(2):219--226, 2019.

\bibitem{menegaux2019continuous}
Romain Menegaux and Jean-Philippe Vert.
\newblock Continuous embeddings of dna sequencing reads and application to
  metagenomics.
\newblock {\em Journal of Computational Biology}, 26(6):509--518, 2019.

\bibitem{HMP}
Barbara~A Meth{\'e}, Karen~E Nelson, Mihai Pop, Heather~H Creasy, Michelle~G
  Giglio, Curtis Huttenhower, Dirk Gevers, Joseph~F Petrosino, Sahar Abubucker,
  Jonathan~H Badger, et~al.
\newblock A framework for human microbiome research.
\newblock {\em nature}, 486(7402):215, 2012.

\bibitem{muggli2017succinct}
Martin~D Muggli, Alexander Bowe, Noelle~R Noyes, Paul~S Morley, Keith~E Belk,
  Robert Raymond, Travis Gagie, Simon~J Puglisi, and Christina Boucher.
\newblock Succinct colored de bruijn graphs.
\newblock {\em Bioinformatics}, 33(20):3181--3187, 2017.

\bibitem{murphy2018janossy}
Ryan~L Murphy, Balasubramaniam Srinivasan, Vinayak Rao, and Bruno Ribeiro.
\newblock Janossy pooling: Learning deep permutation-invariant functions for
  variable-size inputs.
\newblock {\em arXiv preprint arXiv:1811.01900}, 2018.

\bibitem{mustafa2019dynamic}
Harun Mustafa, Ingo Schilken, Mikhail Karasikov, Carsten Eickhoff, Gunnar
  R{\"a}tsch, and Andr{\'e} Kahles.
\newblock Dynamic compression schemes for graph coloring.
\newblock {\em Bioinformatics}, 35(3):407--414, 2019.

\bibitem{approxmatching_mantis}
Prashant Pandey, Fatemeh Almodaresi, Michael~A Bender, Michael Ferdman, Rob
  Johnson, and Rob Patro.
\newblock Mantis: A fast, small, and exact large-scale sequence-search index.
\newblock {\em Cell systems}, 7(2):201--207, 2018.

\bibitem{deepSNP}
Ryan Poplin, Pi-Chuan Chang, David Alexander, Scott Schwartz, Thomas Colthurst,
  Alexander Ku, Dan Newburger, Jojo Dijamco, Nam Nguyen, Pegah~T Afshar, et~al.
\newblock A universal {SNP} and small-indel variant caller using deep neural
  networks.
\newblock {\em Nature biotechnology}, 36(10):983, 2018.

\bibitem{MetaHIT}
Junjie Qin, Ruiqiang Li, Jeroen Raes, Manimozhiyan Arumugam,
  Kristoffer~Solvsten Burgdorf, Chaysavanh Manichanh, Trine Nielsen, Nicolas
  Pons, Florence Levenez, Takuji Yamada, et~al.
\newblock A human gut microbial gene catalogue established by metagenomic
  sequencing.
\newblock {\em nature}, 464(7285):59, 2010.

\bibitem{quail2012tale}
Michael~A Quail, Miriam Smith, Paul Coupland, Thomas~D Otto, Simon~R Harris,
  Thomas~R Connor, Anna Bertoni, Harold~P Swerdlow, and Yong Gu.
\newblock A tale of three next generation sequencing platforms: comparison of
  ion torrent, pacific biosciences and illumina miseq sequencers.
\newblock {\em BMC genomics}, 13(1):341, 2012.

\bibitem{rojas2019genet}
Mateo Rojas-Carulla, Ilya Tolstikhin, Guillermo Luque, Nicholas Youngblut, Ruth
  Ley, and Bernhard Sch{\"o}lkopf.
\newblock Genet: Deep representations for metagenomics.
\newblock {\em arXiv preprint arXiv:1901.11015}, 2019.

\bibitem{schuler199610}
Gregory~D Schuler, Jonathan~A Epstein, Hitomi Ohkawa, and Jonathan~A Kans.
\newblock Entrez: Molecular biology database and retrieval system.
\newblock In {\em Methods in enzymology}, volume 266, pp. 141--162. Elsevier,
  1996.

\bibitem{segata2012metagenomic}
Nicola Segata, Levi Waldron, Annalisa Ballarini, Vagheesh Narasimhan, Olivier
  Jousson, and Curtis Huttenhower.
\newblock Metagenomic microbial community profiling using unique clade-specific
  marker genes.
\newblock {\em Nature methods}, 9(8):811, 2012.

\bibitem{shoeybi2019megatron}
Mohammad Shoeybi, Mostofa Patwary, Raul Puri, Patrick LeGresley, Jared Casper,
  and Bryan Catanzaro.
\newblock Megatron-lm: Training multi-billion parameter language models using
  gpu model parallelism.
\newblock {\em arXiv preprint arXiv:1909.08053}, 2019.

\bibitem{approxmatching1}
Brad Solomon and Carl Kingsford.
\newblock Fast search of thousands of short-read sequencing experiments.
\newblock {\em Nature biotechnology}, 34(3):300, 2016.

\bibitem{svenstrup2017hash}
Dan~Tito Svenstrup, Jonas Hansen, and Ole Winther.
\newblock Hash embeddings for efficient word representations.
\newblock In {\em Advances in Neural Information Processing Systems}, pp.
  4928--4936, 2017.

\bibitem{szmit2013locality}
Rados{\l}aw Szmit.
\newblock Locality sensitive hashing for similarity search using mapreduce on
  large scale data.
\newblock In {\em Intelligent Information Systems Symposium}, pp. 171--178.
  Springer, 2013.

\bibitem{tully2018reconstruction}
Benjamin~J Tully, Elaina~D Graham, and John~F Heidelberg.
\newblock The reconstruction of 2,631 draft metagenome-assembled genomes from
  the global oceans.
\newblock {\em Scientific data}, 5:170203, 2018.

\bibitem{rRNAbayes}
Qiong Wang, George~M Garrity, James~M Tiedje, and James~R Cole.
\newblock Naive bayesian classifier for rapid assignment of rrna sequences into
  the new bacterial taxonomy.
\newblock {\em Appl. Environ. Microbiol.}, 73(16):5261--5267, 2007.

\bibitem{weinberger2009feature}
Kilian Weinberger, Anirban Dasgupta, John Langford, Alex Smola, and Josh
  Attenberg.
\newblock Feature hashing for large scale multitask learning.
\newblock In {\em Proceedings of the 26th annual international conference on
  machine learning}, pp. 1113--1120, 2009.

\bibitem{sequencingcosts}
Kris~A Wetterstrand.
\newblock {DNA} sequencing costs: data from the {NHGRI} genome sequencing
  program ({GSP}), 2013.

\bibitem{NCBIRef}
David~L Wheeler, Tanya Barrett, Dennis~A Benson, Stephen~H Bryant, Kathi
  Canese, Vyacheslav Chetvernin, Deanna~M Church, Michael DiCuccio, Ron Edgar,
  Scott Federhen, et~al.
\newblock Database resources of the national center for biotechnology
  information.
\newblock {\em Nucleic acids research}, 35(suppl\_1):D5--D12, 2006.

\bibitem{wood2019improved}
Derrick~E Wood, Jennifer Lu, and Ben Langmead.
\newblock Improved metagenomic analysis with kraken 2.
\newblock {\em Genome biology}, 20(1):257, 2019.

\bibitem{wood2014kraken}
Derrick~E Wood and Steven~L Salzberg.
\newblock Kraken: ultrafast metagenomic sequence classification using exact
  alignments.
\newblock {\em Genome biology}, 15(3):R46, 2014.

\bibitem{deepsets}
Manzil Zaheer, Satwik Kottur, Siamak Ravanbakhsh, Barnabas Poczos, Ruslan~R
  Salakhutdinov, and Alexander~J Smola.
\newblock Deep sets.
\newblock In {\em Advances in neural information processing systems}, pp.
  3391--3401, 2017.

\bibitem{DeepGenomicsPrimer}
James Zou, Mikael Huss, Abubakar Abid, Pejman Mohammadi, Ali Torkamani, and
  Amalio Telenti.
\newblock A primer on deep learning in genomics.
\newblock {\em Nature genetics}, p.~1, 2018.

\end{thebibliography}

\clearpage
\newpage

\appendix
\onecolumn
\setcounter{table}{0}
\renewcommand{\thetable}{S\arabic{table}}
\setcounter{figure}{0}
\renewcommand{\thefigure}{S\arabic{figure}}

\section{Supplementary material}
\label{app:supp_results}

To train and test our models, we have downloaded $3\,332$ genomes from the NCBI RefSeq database~\cite{NCBIRef}. The full list of accession numbers for the genomes used in our data set can be found in our GitHub repository (\url{https://github.com/ag14774/META2}).


\begin{table}[h]
\centering
\caption{Number of taxa per rank in our data set. The selected accession numbers are a subset of the data set used by~\cite{rojas2019genet}. See subsection~\ref{sec:datagen}}
\label{tab:taxons}
\begin{tabular}{@{}lc@{}}
\toprule
\textbf{Rank} & \multicolumn{1}{l}{\textbf{\# of taxa}} \\ \midrule
Phylum & 37 \\
Class & 86 \\
Order & 184 \\
Family & 372 \\
Genus & 831 \\
Species & 1862 \\ \bottomrule
\end{tabular}
\end{table}

\begin{figure}[h]
  \centering
  \includegraphics[width=0.6\linewidth]{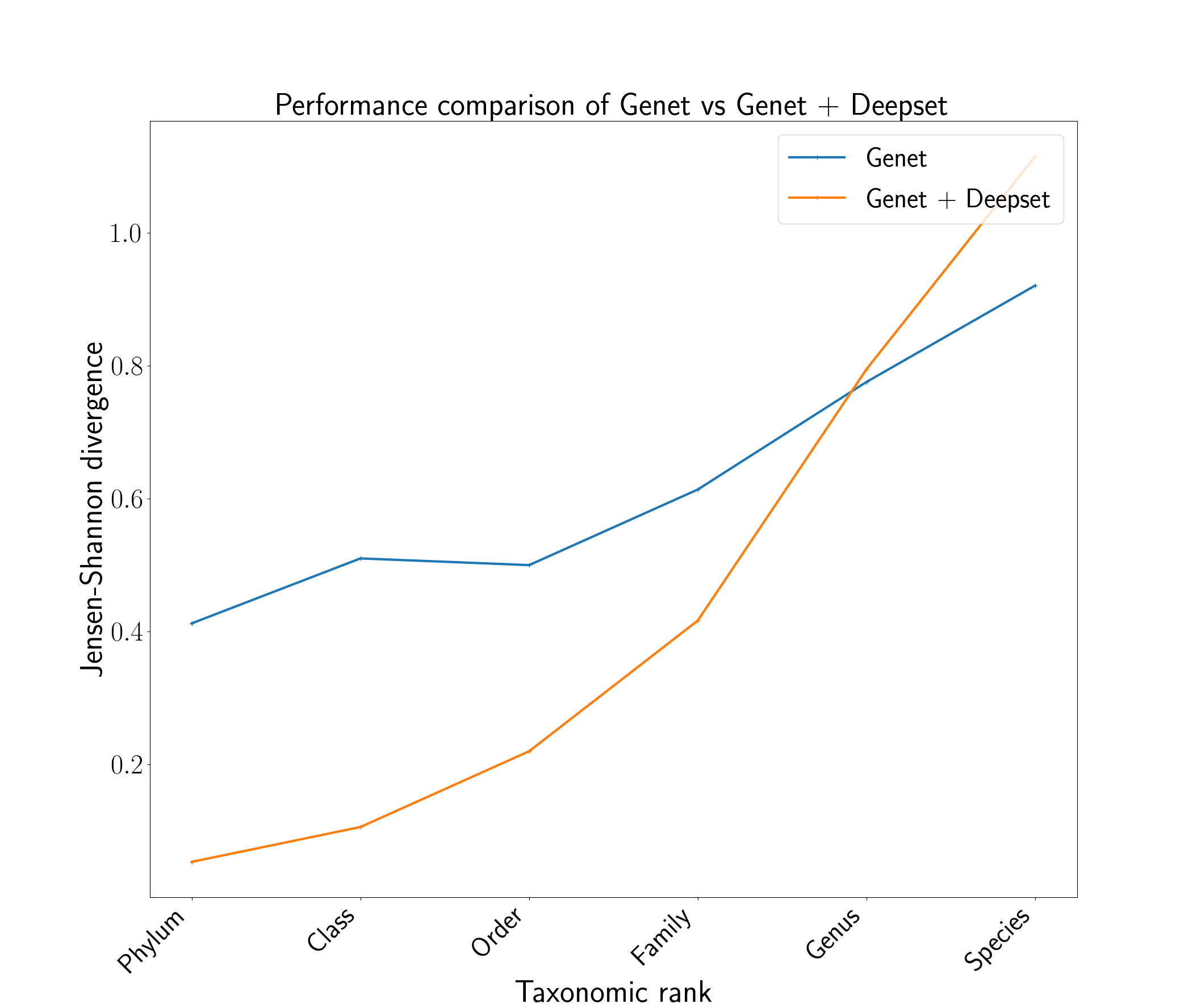}
  \caption{Performance comparison of \textit{GeNet} vs \textit{GeNet + Deepset}. \textit{GeNet + Deepset} achieves superior performance on taxonomic ranks upto \textit{Family}.}
  \label{fig:genetvsds}
\end{figure}

Figure~\ref{fig:genetvsds} shows the performance comparison between our best performing MIL pooling method \textit{GeNet + Deepset} vs standard \textit{GeNet}. It is clear that MIL pooling provides significant improvements at the higher taxonomic ranks.

In addition to training on error-free reads, we attempt to train on noisy NovaSeq reads to determine whether this approach provides any benefit. The JS-divergence achieved by all models when trained on NovaSeq type reads is shown in Table~\ref{tab:js_novaseq} while a comparison of our best performing MIL model, \textit{GeNet + Deepset}, and \textit{GeNet} is depicted in Figure~\ref{fig:genetvsds}.

\begin{table}[h]
\centering
\caption{JS-divergence for all models trained on reads with Illuina NovaSeq noise. Our MIL models achieve superior performance compared to all deep learning models and comparable performance to mapping-based methods at higher taxonomic ranks up to \textit{Family}. No significant improvements occur by directly training on NovaSeq noisy reads compared to error-free reads.}
\label{tab:js_novaseq}
\begin{tabular}{@{}lccc@{}}
\toprule
 & Phylum & Family & Species \\ \midrule
GeNet~\cite{rojas2019genet} & $0.466 \pm 0.062$ & $0.633 \pm 0.064$ & $0.912 \pm 0.057$ \\
EmbedPool~\cite{deepmicrobes} & $0.135 \pm 0.0319$ & $0.478 \pm 0.080$ & $0.733 \pm 0.064$ \\ \midrule
GeNet + Deepset (ours) & $0.053 \pm 0.028$ & $0.417 \pm 0.131$ & $1.115 \pm 0.253$ \\
GeNet + Attention (ours) & $0.062 \pm 0.035$ & $0.462 \pm 0.139$ & $1.135 \pm 0.253$ \\
Embedpool + Deepset (ours) & $0.067 \pm 0.041$ & $0.540 \pm 0.112$ & $1.101 \pm 0.228$ \\
Embedpool + Attention (ours) & $0.066 \pm 0.040$ & $0.539 \pm 0.112$ & $1.107 \pm 0.247$ \\
Hash-Embedpool + Deepset (ours) & $0.060 \pm 0.041$ & $0.539 \pm 0.113$ & $1.202 \pm 0.204$ \\
Hash-Embedpool + Attention (ours) & $0.060 \pm 0.041$ & $0.538 \pm 0.114$ & $1.208 \pm 0.213$ \\ \midrule
Kraken 200~MB & $0.037 \pm 0.021$ & $0.373 \pm 0.054$ & $0.642 \pm 0.078$ \\
Kraken 500~MB & $0.015 \pm 0.008$ & $0.188 \pm 0.049$ & $0.377 \pm 0.061$ \\
Kraken 8~GB & $0.012 \pm 0.014$ & $0.129 \pm 0.038$ & $0.333 \pm 0.094$ \\
Kraken & $0.007 \pm 0.003$ & $0.098 \pm 0.013$ & $0.237 \pm 0.056$ \\
Centrifuge & $0.00007 \pm 0.000001$ & $0.0014 \pm 0.0005$ & $0.030 \pm 0.020$ \\ \bottomrule
\end{tabular}
\end{table}


\clearpage
\section{Hyperparameter grid for the trained models.}
\label{app:hyper}

To train our models, we performed random search over the following hyperparameter grid:

\begin{table}[h]
\centering
\caption{Hyperparameter grid used for training our models}
\label{tab:hyper}
\begin{tabular}{@{}ll@{}}
\toprule
\multicolumn{2}{c}{\textbf{General parameters for single read models}}        \\ \midrule
Batch Size                        & 64, 128, 256, 512, 1024, 2048  \\ \midrule
\multicolumn{2}{c}{\textbf{General parameters for MIL models}}                \\ \midrule
Bag Size                          & 64, 128, 512, 1024, 2048       \\
Batch Size                        & 1, 2, 4, 8                     \\ \midrule
\multicolumn{2}{c}{\textbf{GeNet}}                                 \\ \midrule
Output size of ResNet             & 128, 256, 512, 1024            \\
Use GeNet initialization scheme   & True, False                    \\
BatchNorm running statistics      & True, False                    \\ 
Optimizer                         & Adam, SGD                      \\
Learning rate                     & 0.001, 0.0005, 1.0 (for SGD)   \\
Nesterov momentum (SGD only)      & 0.0, 0.9, 0.99                 \\ \midrule
\multicolumn{2}{c}{\textbf{EmbedPool}}                             \\ \midrule
Size of MLP hidden layer          & 1000, 3000                     \\
Optimizer                         & Adam, RMSprop, SGD             \\
Nesterov momentum (SGD only)      & 0.0, 0.5, 0.9, 0.99            \\
Learning rate                     & 0.001, 0.0005                  \\ \midrule
\multicolumn{2}{c}{\textbf{Deepset pooling layer}}                 \\ \midrule
Deepset $\rho$ hidden layer size  & 128, 256, 1024                 \\
Deepset output size               & 128, 1024                      \\
Dropout before $\rho$ network     & 0.0, 0.2, 0.5, 0.8             \\
Deepset activation                & ReLU, Tanh, ELU                \\ \midrule
\multicolumn{2}{c}{\textbf{Attention pooling layer}}               \\ \midrule
Hidden layer size                 & 128, 256, 512, 1024            \\
Gated attention                   & False, True                    \\
Attention rows                    & 1, 10, 30, 60                  \\ \midrule
\multicolumn{2}{c}{\textbf{MinHash LSH}}               \\ \midrule
LSH $l$ parameter                 & 3, 5, 7            \\ \midrule
\multicolumn{2}{c}{\textbf{Hash Embedding}}               \\ \midrule
Number of hashes $q$                 & 2, 3            \\ \bottomrule
\end{tabular}
\end{table}

\end{document}